\documentclass{ifacconf}

\usepackage{natbib}
\usepackage{amssymb,amsmath,amsfonts,mathtools,bbm}
\usepackage{arydshln}

\usepackage{graphicx,psfrag}
\usepackage{enumitem}

\usepackage[monochrome]{color}

\DeclareMathOperator{\diag}{diag}
\DeclareMathOperator{\bdiag}{blockdiag}

\DeclareMathOperator{\rank}{rank}
\DeclareMathOperator{\Cov}{Cov}
\DeclareMathOperator{\tr}{tr}
\DeclareMathOperator{\spec}{spec}
\DeclareMathOperator{\Alg}{amult}

\newcommand{\norm}[1]{\ensuremath{\left\| #1 \right\|}}

\newcommand{\calA}{\ensuremath{\mathcal{A}}}

\newcommand{\calF}{\ensuremath{\mathcal{F}}}

\newcommand{\calN}{\ensuremath{\mathcal{N}}}
\newcommand{\calO}{\ensuremath{\mathcal{O}}}
\newcommand{\calT}{\ensuremath{\mathcal{T}}}
\newcommand{\calW}{\ensuremath{\mathcal{W}}}

\newcommand{\R}{\ensuremath{\mathbb R}}

\newcommand{\N}{\ensuremath{\mathbb N}}
\DeclareMathOperator{\E}{\mathbb{E}}

\newcommand{\lmin}{\ensuremath{\lambda}_{\textrm{min}}}
\newcommand{\lmax}{\ensuremath{\lambda}_{\textrm{max}}}

\newcommand{\bse}{\begin{subequations}}
	\newcommand{\ese}{\end{subequations}}
\def\be{\begin{equation}}
\def\ee{\end{equation}}

\newcommand{\bbm}{\begin{bmatrix}}
	\newcommand{\ebm}{\end{bmatrix}}

\renewcommand{\qed}{\hfill\blacksquare}
\newcommand{\qedwhite}{\hfill \ensuremath{\Box}}

\usepackage[prependcaption,colorinlistoftodos]{todonotes}

\newtheorem{problem}[thm]{Problem}

\newtheorem{assumption}{Standing Assumption}

\newcommand{\blue}[1]{\textcolor{blue}{#1}}

\newcommand{\red}[1]{\textcolor{red}{#1}}

\begin{document}
\begin{frontmatter}

\title{Gaussian Mechanism Design for Prescribed Privacy Sets in Data Releasing Systems} 

\author[First]{Teimour Hosseinalizadeh} 
\author[First]{Nima Monshizadeh}

\address[First]{Engineering and Technology
	Institute Groningen, University of Groningen, $9747$ AG Groningen ,The Netherlands (e-mail: \{t.hosseinalizadeh, n.monshizadeh\}@rug.nl).}

\begin{abstract}
The data transmitted by cyber-physical systems can be intercepted and exploited by malicious individuals to infer privacy-sensitive information regarding the physical system. 
This motivates us to study the problem of preserving privacy in data releasing of linear dynamical system using stochastic perturbation. \blue{In this study, the privacy sensitive quantity is the initial state value of the system. For protecting its privacy, we directly design the covariance matrix of a Gaussian output noise to achieve a prescribed uncertainty set in the form of hyper-ellipsoids.}
This is done by correlated noise and through a convex optimization problem by considering the utility of released signals. 
\blue{Compared to other available methods, our proposed technique for designing the Gaussian output noise provides enhanced flexibility for system designers. }
As a case study, the results are applied to a heating ventilation and air conditioning system.
\end{abstract}


\end{frontmatter}

\section{introduction}\label{sec:introduction}
Cyber-Physical Systems (CPSs) such as smart grids, intelligent transportation, and smart buildings provide better scalability, fault tolerance, and resource sharing compared to traditional systems. These come at the expense of sharing data and possibly losing privacy in the society.  
   A case in point is the advent of smart electricity meters where they provide benefits such as
     1) giving grid operators better insight of the grid, 2) reducing the cost of taking meter readings and 3) providing consumers with detailed data to help them in reducing their consumption. In its simple form, however, meter readings even at fifteen minute intervals provide a detailed view into a consumer's
personal life. 
 Initial proposal of the law for smart meters in some countries did
not consider consumers' privacy, and was rejected.
 The passed amendment removed the obligation to have smart
meters and allowed people to switch them off administratively; see \cite{van2019smart} for a detailed study.
\par \textit{Related studies.}
For preserving privacy in dynamical 
systems, we make a distinction between preserving privacy in \textit{computation} and \textit{data releasing systems}. 
\blue{Cryptography based methods and approaches from system theory have been studied for private computation in: optimization by \cite{sultangazin2020symmetries} and \cite{Alexandru2020a}, control over the cloud by \cite{kim2022dynamic} and \cite{murguia2020secure}, multiagent systems by \cite{altafini2020system}, \cite{darup2018encrypted}, \cite{monshizadeh2019plausible}, and \cite{hosseinalizadeh2022private}; while studies by  \cite{lu2020privacy} and \cite{tsiamis2017state} are related to preserving privacy in data releasing systems.}
The common feature for cryptography based methods is that while
 they offer strong privacy guarantees and the result of the computation is correct, they are heavy in computation and communication overhead and thus less appealing for CPSs' applications.
On the other hand, methods from system theory while (generally) introduce no errors they offer weaker privacy guarantees.
 \par Stochastic privacy-preserving policies provide solutions for both privacy in computation and data releasing systems (data bases). The most common approach in this case is differential privacy where it adds random noises into each individual’s data such that the statistics of privacy-preserving
outputs do not change by varying the data of
individual; see \cite{dwork2014algorithmic}. Some of the studies using differential privacy in control systems are: private filtering  by \cite{le2013differentially}, average consensus \cite{nozari2017differentially}, distributed optimization \cite{han2016differentially}, \cite{hale2017cloud} and its relation to systems' properties \cite{kawano2020design}. 
Moreover, various measures from information theory have been used to quantify the privacy in dynamical systems (the distribution and the amount of noise) such as Fisher information by \cite{farokhi2019ensuring}, mutual information in  \cite{tanaka2016semidefinite} and  \cite{murguia2021privacy}, and differential entropy in \cite{hayati2021finite}. 
\par \textit{Contribution.} Our problem of interest belongs to preserving the privacy in data releasing systems. We consider the case where the system input and output trajectories are transmitted through a public channel to another party for further processing such as monitoring, safety, or control design while an optimal adversary using public data is interested in determining the state trajectories.
The main contribution of this study is to design a Gaussian \blue{output} perturbation which guarantees a prescribed confusion set for the initial condition of a linear system.
\red{The prescribed confusion set can be shaped to value highly privacy-sensitive state components in the Gaussian mechanism design, and practically ignore privacy-insensitive ones.
This is advantageous in systems where not all state variables have similar importance in view of privacy. }
We prove that any confusion set described by hyper-ellipsoids can be obtained for the unbiased adversaries by utilizing correlated Gaussian noise \blue{at output}. 
This treatment is different from other noise based methods in, \cite{le2013differentially} and \cite{murguia2021privacy} where the confusion set is mapped to a scalar and as we show is predetermined by system dynamics.
The problem for finding an uncorrelated Gaussian \blue{output} noise does not always accept a solution and hence an approximation is provided. 
\par \textit{Notation.} 
The set of positive and nonnegative integers and (positive) real numbers are denoted by $\N$, $\N_{0}$ and ($\R^{+}$) $\R$, respectively. We denote the identity matrix of size $n$ by $I_n$, the zero matrix of size $n$ by $0_n$, and we drop the index whenever the dimension is clear from the context. For a square matrix $A$, $\tr(A)$ and $\det(A)$ denote its trace and determinant; $A^{\dag}$ denote its Moore-Penrose pseudoinverse, $\spec(A)$ and $\spec_{\neq0}(A)$ are the set of its eigenvalues and nonzero eigenvalues, respectively. We denote the algebraic multiplicity for an eigenvalue $\lambda$ of $A$ by $\Alg_{A}(\lambda)$. By $A \succ0$($\succeq 0$), we mean $A$ is a positive(-semi) definite matrix. By $X \sim \calN_n(\mu, \Sigma)$ we denote the random variable $X$ that has the normal distribution with density function $f(x) = (2\pi)^{-n/2}(\det(\Sigma)^{-1/2})\exp(-\frac{1}{2}(x-\mu)^{\top}\Sigma^{-1}(x-\mu))$, where $\mu \in \R^{n}$ and $\Sigma \succ 0$ are mean and covariance, respectively, and $x$ is a realization of $X$.
\par The rest of the paper is organized as follows: In Section \ref{sec:prob_form}, we formulate the problem of interest; Section \ref{sec:out_put_des} designs the output Gaussian mechanism, in Section \ref{sec:Analytical solution}, we present an optimization by considering the performance of the Gaussian noise, Section \ref{sec:Sim} provides a case study and finally Section \ref{sec:conclusion} concludes the paper. 
\section{Problem formulation}\label{sec:prob_form}
We consider linear dynamical systems described by equations of the form
\be\label{eq:lin_sys}
\begin{aligned}
 x(k+1) &= Ax(k) + Bu(k) \\ y(k) &= Cx(k) + Du(k), \quad k \in \N_{0}, 
\end{aligned}
\ee
with state $x \in \R^{n}$, input $u \in \R^{m}$ and output $y \in \R^{p}$.
For this system, define 
\begin{equation*}
\begin{aligned}
	&U_{K-1} \coloneqq \bbm u^\top(0), u^\top(1), \ldots, u^\top(K-1) \ebm^\top \in \R^{mK}\\
 &Y_{K-1} \coloneqq \bbm y^\top(0), y^\top(1), \ldots, y^\top(K-1) \ebm^\top \in \R^{pK}\\
	& \calT_K \coloneqq \bbm
	D & 0 & \cdots & 0 & 0 \\
	C B & D & \cdots & 0 & 0 \\
	\vdots & \vdots & \ddots & \vdots & \vdots \\
	C A^{K-2} B &C A^{K-3}B & \cdots & D &  0 \\
	C A^{K-1} B & C A^{K-2} B & \cdots & C B & D
	\ebm \\
	& \calO_K \coloneqq \bbm C^\top, (CA)^\top, \ldots, (CA^{K-1})^\top \ebm^\top,
\end{aligned}
\end{equation*}
for some $K \in \N$.
Note that the matrices $U_{K-1}$ and $Y_{K-1}$ corresponds to the $K$-long \blue{input and output} trajectories of the system. The matrix $\mathcal{T}_K$ has a Toeplitz structure and contains the Markov parameters of the system and the matrix $\calO_K$ is the $K$-step observability matrix of the system. The above matrices satisfy
\be\label{eq:data_eq}
Y_{K-1} = \calO_K x_0 + \calT_{K}U_{K-1},
\ee
with $x_0$ denoting the initial state of system  \eqref{eq:lin_sys}.
We consider a scenario where the \blue{input ($U_{K-1}$) and output ($Y_{K-1}$)} trajectories of the system \eqref{eq:lin_sys} are transmitted through a public channel to another party for further processing such as monitoring, safety, or control design.
\par We are interested in the case where state variables or some of the state variables contain privacy-sensitive information. 
From \eqref{eq:lin_sys}, it follows that given the system matrices $(A,B)$, and the input of the system, preserving the privacy of the state variables amounts to preserving the privacy of $x_0$. 
\blue{Furthermore, the initial state $x_0$ for stable systems such as a chemical reactor can include valuable information worthy of protection.}
Hence, we treat $x_0$\footnote{The results can be applied for preserving privacy of $x(l)$ for arbitrary $l \in \N_0$ as long as $x(l)$ can be estimated using a window of length $T$ of input/output data.} as a privacy-sensitive value which should remain hidden from any other party, known as \textit{adversary}. 
The adversary's capabilities are specified in the following assumption.
	\begin{assumption}[Adversary's model]\label{assum:adv}
		An adversary $\calA$ knows the system matrices $\big( A, B, C, D\big)$, the released input/output of the system \eqref{eq:lin_sys}, and the exact distribution of the added noises (to be determined  later). 
	\end{assumption}
This type of adversary is also known as honest-but-curious or passive to distinguish it from an active adversary which can manipulate the system. The passive adversaries eavesdrop on communication channels, use public information, and the side knowledge (Assumption \ref{assum:adv}) to infer privacy-sensitive quantities of the system, i.e., $x_0$ in the current setup. 
\par It is well-known that, if the system is observable, then its initial condition can be reconstructed from sufficiently long input-output data samples. Namely, if $\mathcal{O}_K$ has full column rank, then  {\cite[p. 259]{antsaklis}} 
	\be\label{eq:int_con}
	x_0 = \calW_o^{-1}\calO_K^{\top}(Y_{K-1} - \calT_{K}U_{K-1}), 
	\ee
	where $\calW_o \coloneqq \calO_K^{\top} \calO_K$ is called the \textit{observability gramian}.
 It follows from  \eqref{eq:int_con} that, under the observability assumption, the adversary $\calA$ can uniquely determine the initial condition $x_0$ and consequently the state trajectory $x(k)$ for all $k$.
Therefore, we make the following assumption throughout the paper.
\begin{assumption}\label{assum:sys_obs}
The matrix $\mathcal{O}_K$ has full column rank.
\end{assumption}
As a solution for providing privacy for $x_0$, we first consider perturbing the initial condition $x_0$ itself. 
\subsection{Perturbing the initial state}
Assume the perturbed initial condition to be
\be\label{eq:int_noise}
\tilde{x}_0 \coloneqq x_0 + v,
 \ee
where $v$ is a random variable with normal distribution $v \sim \calN_n(0, \Sigma_v)$ and independent of $x_0$.
The data equation \eqref{eq:data_eq} then modifies to
\be\label{eq:Y_tilde_input}
\tilde{Y}_{K-1} = \calO_K x_0 + \calO_K v + \calT_{K}U_{K-1}.
\ee
The following well-known result based on the Gauss-Markov theorem \cite[p. 97]{kailath2000linear} provides the estimation of $x_0$ using an optimal approach by adversary $\calA$.
\begin{lem}[Privacy by adding noise to $x_0$]\label{pro:prv_input_noise}
	Let the perturbed initial state for the system \eqref{eq:lin_sys} be given by \eqref{eq:int_noise}.	
	 The optimum unbiased linear least-mean-squares estimator of $x_0$ is
	\be\label{eq:x0_noise_input}
	\hat{x}_0 =
	\calW_o^{-1}\calO_K^\top(\tilde{Y}_{K-1} - \calT_{K}U_{K-1}), 
	\ee 
	with $\tilde{Y}_{K-1}$ in \eqref{eq:Y_tilde_input}. The covariance of $\hat{x}_0$ is
	\[
	\Cov(\hat{x}_0) = \E\big[(\hat{x}_0 -x_0)(\hat{x}_0 - x_0)^\top\big] = 
	\Sigma_v.
	\]
\end{lem}
The estimator \eqref{eq:x0_noise_input} is also known as maximum likelihood estimator which for Gaussian noise achieves the Cram\'er–Rao bound, the lowest possible bound that any unbiased estimator can obtain. \blue{With the estimation in \eqref{eq:x0_noise_input} and the knowledge set specified in Assumption \ref{assum:adv}, the adversary can optimally estimate $x(k)$ for any $k \geq 1$.}
\par It follows from Lemma \ref{pro:prv_input_noise} that adding noise directly to the initial state $x_0$ enables the designer to hide the true value of $x_0$ within a prescribed confusion set, characterized by $\Sigma_v$.
Despite this advantage, perturbing the initial state is neither feasible nor desired in real-life processes such as a chemical reactor,  since the method requires the system (e.g., the chemical reactor) to be operated with $\tilde{x}_0$ instead of $x_0$, which often requires physical interventions.  
Motivated by this limitation, we consider instead perturbing the measurements of the system $Y_{K-1}$, which can be implemented in the cyber part of CPSs.

\subsection{Perturbing the output measurements}

As an alternative for adding noise to the initial state $x_0$, we perturb the measurement vector $Y_{K-1}$ as
\be\label{eq:Y_per}
\tilde{Y}_{K-1} = Y_{K-1} + N_{K-1} ,
\ee
where the added noise $N_{K-1}\sim\calN_{pK}(0,\Sigma)$ is independent of $Y_{K-1}$. \blue{While a trusted party can remove $N_{K-1}$ from released signals $\tilde{Y}_{K-1}$ by agreeing with the system designer on the seed of the pseudo-random number generator, an adversary can only optimally estimate the initial condition.} This estimation is presented in the following lemma which its proof is analogous to Lemma \ref{pro:prv_input_noise}, and thus is dropped. 
\begin{lem}[Privacy by adding noise to $Y_{K-1}$]\label{pro:prv_output_noise}
	Let the perturbed measurement vector for the system \eqref{eq:lin_sys} be given by \eqref{eq:Y_per}.
 The optimum unbiased linear least-mean-squares estimator of $x_0$ is
	\[
	\hat{x}_0 =
	\big(\calO_K^ \top \Sigma^{-1} \calO_K \big)^{-1}\calO_K^\top \Sigma^{-1}(\tilde{Y}_{K-1} - \calT_{K}U_{K-1}), 
	\]
	and the covariance of $\hat{x}_0$ is
	\[
	\Cov(\hat{x}_0) = \E\big[(\hat{x}_0 -x_0)(\hat{x}_0 - x_0)^\top\big] = 
	\big(\calO_K^ \top \Sigma^{-1} \calO_K \big)^{-1}.
	\]
\end{lem}
Lemma \ref{pro:prv_output_noise} and the discussion succeeding Lemma \ref{pro:prv_input_noise} motivate us to pose the following question: Whether a Gaussian noise $N_{K-1}$ can be found for the output mechanism \eqref{eq:Y_per} such that the optimal adversary in Lemma \ref{pro:prv_output_noise} encounters a prescribed confusion set $\Sigma_v$ for $x_0$?
More formally, we state the following problem: 
\setcounter{thm}{0}
\begin{problem}\label{prb:prob_1}
Find the covariance matrix $\Sigma \succ 0$ for the Gaussian mechanism in \eqref{eq:Y_per} such that 
\be\label{eq:des_prob}
\big(\calO_K^ \top \Sigma^{-1} \calO_K \big)^{-1} = \Sigma_v,
\ee
for a given $\Sigma_v \succ 0$. $\qedwhite$
\end{problem}
\setcounter{thm}{2}
\red{The prescribed confusion set $\Sigma_v$ can be shaped to value highly privacy-sensitive state components in the Gaussian mechanism design, and practically ignore privacy-insensitive ones.} Working with the full matrix $\Sigma_v$ rather than $\sigma I_n$, with $\sigma \in \R^{+}$, is particularly advantageous in systems where not all state variables have similar importance in view of privacy.

\par The \textit{statistical} interpretation for the confusion set originates from the notion of confidence set (region) of an estimation; see \cite{adkins1990improved}.
 A $(1- \alpha)$ confidence set for a parameter ${x_{0}} \in \R^n$ is a set $Q$ such that 
\be\label{eq:confidence}
[\mathbb{P}({x_{0}} \in Q)]\geq 1 - \alpha,
\ee 
where $\alpha = 0.05$ is a common choice.
For a point estimator with a normal distribution, i.e., $\hat{x}_0 \sim \calN_{n}({x_{0}}, \Sigma_v)$ a common confidence set $Q$ is the so-called confidence ellipsoid characterized by $\Sigma_v$ and is defined as
\be\label{eq:ellipsoid}
Q({\Sigma_v}, \gamma) \coloneqq
\{{x_{0}}|
({x_{0}} - \hat{x}_0)^\top \Sigma_v^{-1} ({x_{0}} - \hat{x}_0) \leq \gamma
\},
\ee
where $\gamma$ is a function of $n$ and $\alpha$.
This expression \eqref{eq:confidence} implies that the true values of ${x_{0}}$
are $100(1-\alpha)$ percent of the time in repeated samples within the ellipsoid given in \eqref{eq:ellipsoid}. \blue{ Notice, by choosing nondiagonal $\Sigma_v$ we can shape the orientation of the resulted ellipsoids in \eqref{eq:ellipsoid}.}
\section{Output Gaussian Mechanism}\label{sec:out_put_des}
In this section, we solve Problem \ref{prb:prob_1} by finding the set of positive definite matrices $\Sigma$ satisfying \eqref{eq:des_prob}. 
We draw on the following lemmas in answering the design problem in \eqref{eq:des_prob}. Lemma \ref{lem:matrix_equation_1} provides us with the solutions of the matrix equations of the form \eqref{eq:des_prob}, and Lemma \ref{lem:detectability} is a classical result on detectability of linear system, which we include to make the paper self-contained.

\begin{lem}\cite[Theorem 13.27]{laub2005matrix}\label{lem:matrix_equation_1}
	Let $E\in \R^{m\times n}$, $F\in \R^{p\times q}$, and $G\in \R^{m\times q}$. Then the equation
	\be\label{eq:GMeqution_1}
	EXF=G,
	\ee
	has a solution $X \in \R^{n \times p}$ if and only if 
	\be \label{eq:GMuniqness}
	EE^{\dag}GF^{\dag}F = G,
	\ee 
	 in which case the general solution of \eqref{eq:GMeqution_1} is of the form
	\be\label{eq:GMsolution}
	X = E^{\dag}GF^{\dag} + R - E^{\dag}ERFF^{\dag},	
	\ee
	where $R\in \R^{n\times p}$ is arbitrary. A solution, provided that it exists, is unique if $E$ has full
	column rank and $F$ has full row rank, i.e., $E^{\dag}E = I$ and  $FF^{\dag} = I$.
\end{lem}
\begin{lem}\cite[p.192]{hespanha2018linear}\label{lem:detectability}
	For the linear system 
	\be\label{eq:fic_sys}
	 x({k+1}) = Mx({k}), \quad
	y({k}) = Nx({k}),
	\ee
	with $x({k}) \in \R^{n}$ and $y(k) \in \R^{p}$, the following statements are equivalent:
	\begin{enumerate}
		\item The pair $(M, N)$ is detectable.
		\item For  $|\lambda|\geq1$, with $\lambda \in \spec(M)$, we have
			\be\label{eq:det_second}
				\rank \bbm M-\lambda I \\ N \ebm = n.
			\ee
		\item There exists $P \succ 0$ such that $M^{\top}PM - P - N^{\top}N \prec 0$.
	\end{enumerate}
\end{lem}
The following theorem addresses Problem \ref{prb:prob_1} by providing results on existence and uniqueness of positive definite solutions $\Sigma$ to \eqref{eq:des_prob}.
\begin{thm}\label{thrm:sigma_exists}
	Let $\Sigma_v \succ 0$ be the prescribed confusion set for $x_0$ in Problem \ref{prb:prob_1}.
	Consider the matrix equation
	\be\label{eq:GMX_condition}
	(\calO_K^ \top X \calO_K)^{-1}= \Sigma_v,
	\ee
	and the set 
	\be\label{eq:set_X}
	S_{X} \coloneqq \{ X\in \R^{pK\times pK}| \;\eqref{eq:GMX_condition}\,\,\text{holds} \}.
	\ee
 Then, the following statements hold.
	\begin{enumerate}[label=(\emph{\roman*})]
		\item \label{item_X-exist} The set $S_X$ is nonempty and given by
		\be\label{eq:set_X_R}
			S_{X} = \{ N^{\top}N + R - MRM|\,\, R \in\R^{pK\times pK} \},
		\ee
		where  
		\be \label{eq:M_N_pair}
		M \coloneqq \calO_K\calW_o^{-1} \calO_K^ \top, \quad N \coloneqq \Sigma_v^{-1/2}\calW_o^{-1} \calO_K^ \top.
		\ee
		\item \label{item_Set_X_positive}  	The set 
		\be\label{eq:set_X_positive}
		S_X^+ \coloneqq \{X\in S_X|\,\,X\succ 0 \},
		\ee
		is nonempty.
		\item \label{item_Set_X_single} If $pK=n$, then the set $S_X^+$ is singleton and is given by $S_X^+ = \{ N^{\top}N \}$.$\qedwhite$
	\end{enumerate}
\end{thm}
\textit{Proof.}
	 	\textit{Statement \ref{item_X-exist}:}
We resort to Lemma \ref{lem:matrix_equation_1} to show the existence of a matrix $X$ satisfying \eqref{eq:GMX_condition}. By choosing $E = \calO_K^ \top$, $F = \calO_K$, and $G = \Sigma_v^{-1}$, the condition in \eqref{eq:GMuniqness} is verified as %
	\begin{align*}
	 (\calO_K^ \top)(\calO_K \calW_o^{-1})(\Sigma_v^{-1})(\calW_o^{-1} \calO_K^ \top)(\calO_K) = \Sigma_v^{-1},
	\end{align*}
	where we used the facts that $E^{\dag} = \calO_K \calW_o^{-1}$, $F^{\dag} = \calW_o^{-1} \calO_K^ \top $, and $\calW_o =  \calO_K^ \top \calO_K$. This proves the existence claim in Statement \ref{item_X-exist}.

	It follows from \eqref{eq:GMsolution} in Lemma \ref{lem:matrix_equation_1} that
	every $X$ satisfying \eqref{eq:GMX_condition} is given by 
	\[
		X = \calO_K \calW_o^{-1}\Sigma_v^{-1}\calW_o^{-1} \calO_K^ \top + R- \calO_K \calW_o^{-1}\calO_K^ \top R \calO_K\calW_o^{-1} \calO_K^ \top.
	\]	
By substituting $M$ and $N$ from \eqref{eq:M_N_pair} in the above, we obtain the set $S_X$ in \eqref{eq:set_X_R}.

\textit{Statement \ref{item_Set_X_positive}:}
	First, we observe that the set $S_{X}^+$ requires the matrix $R$ in \eqref{eq:set_X_R} to be symmetric, i.e. $R = R^{\top}$. Then, $S_X^+$ is nonempty if and only if there exists a solution $R=R^\top$ to the following linear matrix inequality:
	\be\label{eq:LMI_sigma}
	N^{\top}N + R - MRM \succ 0.
	\ee
 \par To prove feasibility of the LMI \eqref{eq:LMI_sigma}, consider a fictitious  linear system \eqref{eq:fic_sys} given by the pair $(M, N)$ in \eqref{eq:M_N_pair}. The main idea is to show that the pair $(M, N)$ is detectable, and thus from Lemma \ref{lem:detectability}, the LMI \eqref{eq:LMI_sigma} which can be seen as a Lyapunov inequality for detectability, equivalently holds. For the matrix $M$, we have
	  \be \label{eq:eig_M}
	\begin{aligned}
	 \spec_{\neq0}(M)  & = \spec_{\neq0}(\calO_K\calW_o^{-1} \calO_K^ \top) \\& =  \spec_{\neq0}(\calW_o^{-1} \calO_K^ \top\calO_K)  =  \spec(I_n),
	\end{aligned}
	\ee
where the second equality  follows from the fact that for two arbitrary matrices $A \in \R^{q\times r}$ and $B\in \R^{r \times q}$  the nonzero eigenvalues of $AB$ and $BA$ are the same, with the same algebraic multiplicities \cite[p. 214]{garcia2017second}. Hence, $\spec(M)=\{0,1\}$ 
	 with $\Alg_{M}(0)= pK-n$ and $\Alg_{M}(1) = n$.
	  \par Next, we draw on the second statement in Lemma \ref{lem:detectability} for the detectability of the pair $(M, N)$. 
	Noting that $\lambda=1$ is the only (marginally) unstable eigenvalue, the rank condition in \eqref{eq:det_second} gives rise to
	\be \label{eq:MN_detectability}
	\rank \bbm \calO_K\calW_o^{-1} \calO_K^ \top-I_{pK} \\ \Sigma_v^{-1/2}\calW_o^{-1} \calO_K^ \top \ebm = pK.
	\ee
	Assume that there exists $\zeta \in \R^{pK}$ such that 
	\bse
	\be\label{eq:M_condition}
	(\calO_K\calW_o^{-1} \calO_K^ \top - I)\zeta = 0 
	\ee
	\be\label{eq:N_condition}
	\Sigma_v^{-1/2}\calW_o^{-1} \calO_K^ \top \zeta = 0.
	\ee
	\ese
	Since $\Sigma_v \succ 0$, it follows from \eqref{eq:N_condition} that $(\calW_o^{-1} \calO_K^ \top) \zeta = 0$. This, together with \eqref{eq:M_condition} and Standing Assumption \ref{assum:sys_obs},  results in $\zeta = 0$. Hence, we conclude that  \eqref{eq:MN_detectability} holds, and thus the pair $(M, N)$ is detectable. By the third statement of Lemma \ref{lem:detectability}, 
	there exists $R \succ 0$ such that \eqref{eq:LMI_sigma} holds, and consequently $S_X^+$ is nonempty.  
\par \textit{Statement \ref{item_Set_X_single}:}
	It follows from Lemma \ref{lem:matrix_equation_1} that $X$ satisfying \eqref{eq:GMX_condition} is unique when $\calO_K$ has full row rank. Moreover, by Standing Assumption \ref{assum:sys_obs}, we have $pK \geq n$, and thus $X$ is unique if $pK=n$. In this case,
	$R - MRM = 0$ in \eqref{eq:set_X_R} and hence $S_X = \{N^{\top}N\}$.
The proof is complete by noting that $N^\top N \succ 0$ due to Standing Assumption \ref{assum:sys_obs}. 
$\qed$
It follows from Theorem \ref{thrm:sigma_exists} that the solution to \eqref{eq:GMX_condition} is in general not unique.  
In fact, any solution $R=R^\top$ to the LMI in \eqref{eq:LMI_sigma} returns an admissible solution $\Sigma$ to \eqref{eq:des_prob}.  In the next section, we leverage on this degree of freedom to look for solutions that are superior in terms of performance of the system \eqref{eq:lin_sys}.
\section{Optimal Gaussian Mechanism design}\label{sec:Analytical solution}
We provide a performance measure for the Gaussian noise in \eqref{eq:des_prob}, and formulate an optimization problem to derive the best performance for a given confusion set $\Sigma_v$. 
\par To differentiate among the admissible solutions ($\Sigma$) in \eqref{eq:des_prob}, we first need a notion of performance for the system. Noting that the amount of sensor measurements perturbations in \eqref{eq:Y_per} directly affects the {\em utility} of the signal $Y_{K-1}$, %
we define the error resulting from the perturbation $N_{K-1}$ as\footnote{The error signal $Z_{K-1}$ is in fact equal to the noise signal $N_{K-1}$; however, we opt for the former since \eqref{eq:mse_noise_output} can be viewed as a performance metric independent of the adopted perturbation technique.}
\be\label{eq:mse_noise_output}
\E(Z_{K-1} ^\top Z_{K-1}),
\ee
\blue{where $Z_{K-1} \coloneqq \tilde{Y}_{K-1} - {Y}_{K-1}$. By \eqref{eq:mse_noise_output} we measure the average effects of the added noise on $Y_{K-1}$.} The expression \eqref{eq:mse_noise_output} can be rewritten in terms of covariance of the added noise $\Sigma$ as 
\be\label{eq:mse_out_proof}
\begin{aligned}
	\E(Z_{K-1} ^\top Z_{K-1}) &= \E (\tr(Z_{K-1} Z_{K-1} ^\top)) \\ & =   \tr \E (N_{K-1} N_{K-1} ^\top)  = \tr(\Sigma) .
\end{aligned}
\ee
By taking $\tr(\Sigma)$ as our performance metric, we 
propose the following optimization problem in order to find performance-optimal solution $X\succ 0$ to \eqref{eq:GMX_condition}, and thus $\Sigma= X^{-1}$ to \eqref{eq:des_prob}:
\be\label{eq:opt_nonconv}
\begin{aligned}
	\min_{R \in \R^{pK \times pK}, \epsilon \in \R^{+}}  \,& \epsilon  \\
	\text{sub}\text{ject}& \,\, \text{to}  \\
	&\underbrace{ N^{\top}N + R - MRM}_{:=X} \succ  0 \\
	&\tr X^{-1} \leq \epsilon  .
\end{aligned}
\ee
\blue{Observe from the last constraint in optimization \eqref{eq:opt_nonconv} that we are interested in preserving the privacy of $x_0$ with the minimum amount of distortion of the system output $Y_{K-1}$.}
Noting that the feasibility set in optimization \eqref{eq:opt_nonconv} is  nonconvex in decision variable $R$, we derive a convex approximation for it.
To this end, we upper bound $\tr X^{-1}$  as 
\[
	 \tr X^{-1} \leq  (pK)\lmax(X^{-1}) = \frac{pK}{\lmin(X)},
\]
where we used the fact that $\lmax(X^{-1}) =1/\lmin(X)$. Consequently, a sufficient condition for imposing $\tr X^{-1} \leq \epsilon$ in   \eqref{eq:opt_nonconv} is  given by
\[
\frac{pK}{\lmin(X)} \leq \epsilon,
\]
 This can be equivalently rewritten as 
\[
X \succeq \frac{pK}{\epsilon} I.
\]
By defining $\beta \coloneqq \frac{pK}{\epsilon}$, we replace \eqref{eq:opt_nonconv} by the following convex optimization problem: 
\be\label{eq:main_optimization}
\begin{aligned}
	\max_{R \in \R^{pK \times pK}, \beta \in \R^+}  \,& \beta  \\
		\text{sub}\text{ject}& \,\, \text{to}  \\
	& N^{\top}N + R - MRM \succeq \beta I.
\end{aligned}
\ee
In what follows, we provide the optimal value of $\beta$ in the above maximization problem. 
For doing so, we first recap the following algebraic result:  
\begin{lem}\cite[p. 284]{garcia2017second}\label{lem:A_B_com_simul}
Let $\calF \subseteq \R^{n\times n}$ be a nonempty set of matrices. Suppose that each matrix in $\calF$ is real and symmetric. Then $AB = BA$ for all $A$, $B \in \calF$ if and only if there is a real orthogonal $Q \in \R^{n\times n}$ such that $Q^{\top}AQ$ is diagonal for every $A\in \calF$.
\end{lem}

\begin{thm}\label{thrm:optimization_sol}
	Let $\Sigma_v \succ 0$ be the prescribed privacy set for $x_0$ in Problem \ref{prb:prob_1}.
	Consider the convex optimization problem given in \eqref{eq:main_optimization}, with the matrices $M$, and $N$ in \eqref{eq:M_N_pair}. The optimal value of the objective function is $\beta_{\text{opt}} = \lambda_{\min}(\Sigma_v^{-1}\calW_o^{-1})$.$\qedwhite$
\end{thm}
\textit{Proof.}
	Recall from \eqref{eq:eig_M} that $M = \calO_K\calW_o^{-1} \calO_K^ \top $ has $\spec(M)=\{0,1\}$ where $\Alg_{M}(0)=pK-n$ and $\Alg_{M}(1)=n$. Similarly for $N^{\top}N = \calO_K \calW_o^{-1}\Sigma_v^{-1}\calW_o^{-1} \calO_K^ \top$, we have
	  \[ 
	  \spec(N^{\top}N)=\{0\} \cup \spec(\Sigma_v^{-1}\calW_o^{-1}),
	  \]
	 where $\Alg_{N^{\top}N}(0) = pK-n$. 
\par 
Observe that the matrix $M$ commutes with $N^{\top}N$, and thus they are simultaneously diagonalizable by Lemma \ref{lem:A_B_com_simul}. Namely, there 
exists an orthogonal matrix $S \in \R^{pK \times pK}$ such that   
	\be\label{eq:M_NNT_diag}
	\begin{aligned}
	M =   S\left[
		\begin{array}{c|c}
			I_n & 0 \\
			\hline
			0 & 0_{pK-n}
		\end{array}
		\right]
		S^{\top},\quad  
		 N^{\top}N =S\left[
		\begin{array}{c|c}
			\Lambda & 0 \\
			\hline
			0 & 0_{pK-n}
		\end{array}
		\right]S^{\top}
	\end{aligned},
	\ee
	where $\Lambda = \diag (\lambda_{1},\lambda_{2}, \ldots, \lambda_{n})$ denote the nonzero eigenvalues of $N^\top N$ arranged in a non-increasing order.
	Notice that $\ker M = \ker N^{\top}N$, which allows to write the decomposition in 
	\eqref{eq:M_NNT_diag} corresponding to zero and nonzero eigenvalues. 
	\par Next, 
		partition $S$ and $R$ consistently as
	\[ S =
	\left[
	\begin{array}{c|c}
	S_{11} & S_{12} \\
	\hline
	S_{21} & S_{22}
	\end{array}
	\right], \quad R = \left[
	\begin{array}{c|c}
	R_{11} & R_{12} 
	\\
	\hline
	{R}_{12}^{\top}\textcolor{white}{\hat{I}} & R_{22}
	\end{array}
	\right],
	\]
where $S_{11}, R_{11}\in \R^{n\times n}$. Next, we apply the congruence transformation associated with $S$ to the constraint in \eqref{eq:main_optimization}:
	\begin{equation*}
	\begin{aligned}
		 (S^{\top}MS) (S^{\top}RS) (S^{\top}MS) &- S^{\top}RS  \\ &- S^{\top} (N^{\top}N - \beta I) S  \preceq 0,
	\end{aligned}
	\end{equation*}
	which in block partitioned form is
	\begin{align*}
		\left[
		\begin{array}{c|c}
			I_n & 0 \\
			\hline
			0 & 0_{pK-n}
		\end{array}
		\right]	 \left[
		\begin{array}{c|c}
			\hat{R}_{11} & \hat{R}_{12} \\
			\hline
			\hat{R}_{12}^{\top} & \hat{R}_{22}
		\end{array}
		\right]		&\left[
		\begin{array}{c|c}
			I_n & 0 \\
			\hline
			0 & 0_{pK-n}
		\end{array}
		\right]   -  \left[
		\begin{array}{c|c}
			\hat{R}_{11} & \hat{R}_{12} \\
			\hline
			\hat{R}_{12}^{\top} & \hat{R}_{22}
		\end{array}
		\right] \\  & 
		-\left[
		\begin{array}{c|c}
			\Lambda - \beta I_{n} & 0 \\
			\hline
			0 & -\beta I_{pK-n}
		\end{array}
		\right] \preceq 0,
	\end{align*}
 where 
 \[
S^\top RS \eqqcolon \hat R= \left[
		\begin{array}{c|c}
			\hat{R}_{11} & \hat{R}_{12} \\
			\hline
			\hat{R}_{12}^{\top} & \hat{R}_{22}
		\end{array}
		\right].
\]
The above inequality further simplifies to
	\be\label{eq:LMI_block_feas}
	\begin{aligned}
		 \left[
		\begin{array}{c|c}
			\begin{array}{ccc}
				\lambda_{1} -\beta &  & 0\\
				& \ddots & \\
				0& & \lambda_{n} - \beta
			\end{array} & \hat{R}_{12} \\
			\hline
			\hat{R}_{12}^{\top} & \hat{R}_{22}-\beta I_{pK-n}
		\end{array}
		\right] \succeq 0.
	\end{aligned}
	\ee
	A necessary condition for \eqref{eq:LMI_block_feas} is
	\[
	\beta \leq \lambda_{n},
	\]
which implies that $\beta_{\text{opt}} \leq \lambda_{n}$. 
Moreover, by choosing 
$\hat{R}_{12} = 0$ and $\hat{R}_{22}$ such that $\hat{R}_{22} - \beta I \succeq 0$, we conclude that 
the choice $\beta= \lambda_{n}$ is a feasible solution to \eqref{eq:LMI_block_feas}. Now, since $R = S\hat{R}S^{\top}$, we find that there exists $R$ such that the LMI \eqref{eq:main_optimization} holds for $\beta= \lambda_n$, thereby proving $\beta_{\text{opt}} = \lambda_n$. The proof is complete by noting $\lambda_n = \lambda_{\min}(\Sigma_v^{-1}\calW_o^{-1})$.
$\qed$

By Theorem \ref{thrm:sigma_exists}, the covariance matrix can be designed as $\Sigma^{-1} = N^{\top}N + R - MRM $ where
$R=R^\top$ is any solution to the LMI in \eqref{eq:main_optimization} with $\
\beta=\beta_{\rm opt}$.

Recalling $\beta = \frac{pK}{\epsilon}$, the optimal performance is given by
\be\label{eq:epsilon_condition}
\epsilon_{\text{opt}} = pK\lambda_{\max}(\Sigma_v\calW_o).
\ee
It follows from \eqref{eq:epsilon_condition} that the error caused by the output perturbation \eqref{eq:Y_per} is proportional to the total time steps $K$ and the number of outputs of the system $p$.
Moreover, we observe that both observability degree and the desired privacy guarantees contribute to the optimal performance; namely $\epsilon_{\rm opt}$ is proportional to the spectral norm of the product of the observability gramian and the prescribed privacy set $\Sigma_v$.

\par We close this section by a few remarks on the proposed results and their potential extensions. 

\begin{rem}[Structured output mechanism]\label{rem:Str_Out_Mec}
An extension for the design mechanism \eqref{eq:des_prob} is to impose a specific structure on the covariance matrix $\Sigma$ of the output mechanism in \eqref{eq:Y_per}.
A case of particular interest is given by
the block diagonal structure
	\[
	\Sigma_{\text{blk}} = \bdiag\big(\Sigma_1, \Sigma_2, \ldots, \Sigma_K \big) %
	\]
	with $\Sigma_{k} \in \R^{p\times p}$.
The interest in the block diagonal structure stems from the fact that the output perturbation in \eqref{eq:Y_per} can be then implemented by using uncorrelated noise signals, which is favorable in online applications.
Working with the block diagonal structure in \eqref{eq:des_prob} modifies Problem \ref{prb:prob_1} to: find $\Sigma_{blk} \succ 0$ for Gaussian mechanism in \eqref{eq:Y_per} such that 
	\[
	\big(\calO_K^ \top \Sigma_{blk}^{-1} \calO_K \big)^{-1} = \Sigma_v,
	\]
	for a prescribed $\Sigma_v \succ 0$. By following analogous steps as before, we obtain a counterpart of \eqref{eq:opt_nonconv} as 
	\be\label{eq:opt_Sig_str}
	\begin{aligned}
		\min_{\Sigma_1, \ldots, \Sigma_K \succ 0, \epsilon_{blk} \in \R^{+}}  \,& \epsilon_{blk}  \\
			\text{sub}\text{ject}& \,\, \text{to}  \\
		& \calO_K^ \top \Sigma_{\text{blk}}^{-1}\calO_K  = \Sigma_v^{-1} \\
		& \tr\Sigma_{blk} \leq \epsilon_{blk} .
	\end{aligned}
	\ee
Unlike \eqref{eq:opt_nonconv}, feasibility of \eqref{eq:opt_Sig_str} depends on the choice of $\Sigma_v$. 
A possible remedy to overcome this challenge is to relax the equality constraint and replacing it by a solution $\Sigma_{\text{blk}}$ that (approximately) results in a prescribed confusion set $\Sigma_v$; namely, 
	\[
	\norm{\calO_K^ \top \Sigma_{\text{blk}}^{-1}\calO_K  - \Sigma_v^{-1}}^{2}_{F} \leq e_{blk},
	\]
	where $e_{blk} > 0$ determines the accuracy level of the solution and  
 $\norm{(\cdot)}_{F}$ denotes the Frobenius norm.
\end{rem}

\begin{rem}[Confusion set in differential privacy]
	In order to preserve the privacy of $x_0$, the system designer can use differential privacy to find the covariance $\Sigma$ for the Gaussian noise in \eqref{eq:Y_per}. Generally speaking, this method determines the covariance as $\Sigma = \sigma I_{pK}$ with $\sigma \geq f(\epsilon, \delta, c, s)$ where $(\epsilon, \delta)$ are predefined user's parameters, $c$ defines the adjacency metric for $x_0$, and $s$ is the sensitivity of the output vector $Y_{k-1}$ to changes in $x_0$ \cite[Theorem 3]{le2013differentially}. 
	Based on Lemma \ref{pro:prv_output_noise}, the confusion set that results from this choice of covariance is $\Sigma_v = \sigma \calW_o^{-1}$, i.e., a proportion of the inverse of observability gramian $\calW_o$.
	 Therefore, the shape of the confusion set in the differential privacy is predetermined by $\calW_o$, namely the dynamics of the system \eqref{eq:lin_sys}, while in our method we are interested in the case where the confusion set can be shaped by the designer.
\end{rem}
\begin{rem}[Confusion set for $K \to \infty$]\blue{
While we have designed $\Sigma$ in \eqref{eq:des_prob} for a finite time step $K$, 
we need to consider (Schur) stability properties of system \eqref{eq:lin_sys} to design $\Sigma$ when $K \to \infty$. If system \eqref{eq:lin_sys} is Schur stable and for $\Sigma = \sigma I_{pK}$ in \eqref{eq:des_prob}, the confusion set when $K \to \infty$ is $\Sigma_v(\infty) = \sigma \calW_o^{-1}(\infty)$ where $\calW_o(\infty) \succ 0$ is the unique and bounded solution to the Lyapunov equation for observability \cite[p.192]{hespanha2018linear}
$$A^{\top}\calW_o A - A + C^{\top}C = 0.$$
It follows that for stable systems and the Gaussian mechanism $\Sigma = \sigma I_{pK}$ with finite $\sigma \in \R_+$, the adversary faces the confusion set $\sigma \calW_o^{-1}(\infty)$ when $K \to \infty$. On the other hand, if system \eqref{eq:lin_sys} is unstable, $\calW_o(K)$ grows unboundedly as $K\to\infty$. Thus from $\sigma \calW_o^{-1}(\infty)$ it follows that we  should use large amount of noise (higher value of $\sigma$) to preserve privacy of $x_0$. It is worth mentioning that noise to signal ratio is not necessarily increasing for unstable systems.}
\end{rem}

\section{ Comparison and Simulation}\label{sec:Sim}
We provide a privacy preserving method from the literature for comparison purposes, and then present a case study to illustrate the concepts in this paper. 
\subsection{differential entropy measure}
Another approach for privacy preserving in data releasing system is to map the adversary uncertainty set to a scalar metric. To prepare for the case study, we consider differential entropy as the privacy metric. %
 Differential entropy of a random variable $X$ with normal distribution $X\sim\calN_n(\mu, \Sigma)$ is \cite[p. 250]{thomas2006elements}
\[
h(X) = \frac{1}{2}\log\det(\Sigma)+\frac{n}{2}(1+\log(2\pi)),
\]
where $\log$ is base $2$. It is related to the volume of a set defined by the random variable $X$, and essentially the higher value for it indicates the random variable is widely dispersed.  
Consistently, we consider differential entropy of the adversary's best estimation of $\hat{x}_0$ given in Lemma \ref{pro:prv_output_noise} $\big(\calO_K^ \top \Sigma_{\text{de}}^{-1} \calO_K \big)^{-1}$ as the objective function to be maximized. Analogous to \cite{hayati2021finite}, we consider the following optimization problem to obtain the covariance of the output noise in \eqref{eq:Y_per}:
\be\label{eq:opt_diff}
\begin{aligned}
	\max_{\Sigma_{\text{de}} \in \R^{pK \times pK}}  \,& \log\det\big(\calO_K^ \top \Sigma_{\text{de}}^{-1} \calO_K \big)^{-1}  \\
		\text{sub}\text{ject}& \,\, \text{to}  \\
	&\tr\Sigma_{\text{de}} \leq \epsilon_{p}\\
	&\Sigma_{\text{de}} \succ 0,
\end{aligned}
\ee
where $\epsilon_{p}$ is a predefined performance budget, i.e., \blue{an upper bound on the average error caused by introducing output noise (see \eqref{eq:mse_out_proof}).} 
It should be noted that $\log\det\big(\calO_K^ \top \Sigma_{\text{de}}^{-1} \calO_K \big)^{-1}$ is concave in $\Sigma_{\text{de}}$; see \cite[Proposition 10.6.17]{bernstein2018scalar} and hence the optimization problem \eqref{eq:opt_diff} obtains its global maximum. 
\begin{rem}[Comparison with differential entropy]\label{rem:compare_methods}
	We \\highlight the differences and complementarity  of our proposed method (summarized in optimization problem \eqref{eq:main_optimization}) and the approach \eqref{eq:opt_diff}. 
 The optimization problem \eqref{eq:opt_diff} maximizes a privacy metric for a given performance budget. 
 On the other hand, given a prescribed  \textit{confusion set} $\Sigma_v$, the optimization problem \eqref{eq:main_optimization} looks for the best performance that can be obtained for the Gaussian mechanism by a convex program. 
As the designer can shape the confusion set faced by the adversary, the latter approach is particularly useful in scenarios where different state components have different privacy sensitivity.  
\end{rem}

We further illustrate the comparison mentioned in Remark \ref{rem:compare_methods} in the next subsection.
\subsection{Case study}
As an instance of releasing data in linear dynamical system, we consider a Heating, Ventilation, and Air Conditioning (HVAC) system with the following model from \cite{kelman2011bilinear} 
\be\label{eq:HVAC_dynamic}
M\frac{d}{dt}T_z = RT_z + {Q}_{\text{offset}}(t) + c_p{m}_z(t)\circ(T_s - T_z),
\ee
where $M = \diag\big((mc)_1,\ldots, (mc)_n\big)$, $T_z = \bbm T_{z1}, \ldots, T_{zn}\ebm^{\top}$, ${m}_z(t) = \bbm {m}_{z1}(t), \ldots, {m}_{zn}(t)\ebm^{\top} $, $T_s =  \bbm T_{s1}, \ldots, T_{sn}\ebm^{\top} $, $R = \bbm R_{ij}\ebm$, and $\circ$ denotes the Hadamard product. 
\begin{figure}[ht]
	\begin{center}
		\includegraphics[width=0.35\textwidth]{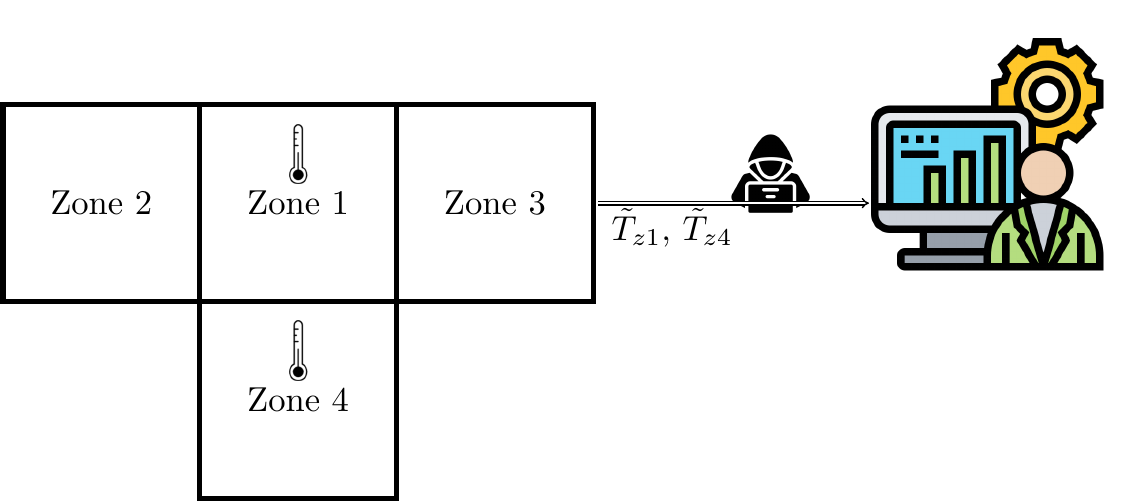}
		\caption{Simplified floor plan of the zones and monitoring systems. The measured temperatures $T_{z1}$ and $T_{z4}$ are perturbed using Gaussian noise and transmitted to another party }\label{fig:zones}
	\end{center}
\end{figure}
\par The system \eqref{eq:HVAC_dynamic} is known as zone temperature dynamic which relates temperate of each zone ($T_{zi}$) in an area with $n$ zones to physical parameters such as heat capacity of air ($c_p$), thermal capacitance of zone $i$ denoted as $(mc)_i$, thermal resistance of heat transfer between zone $i$ and $j$ denoted as $R_{ij}$, mass flow rates ${m}_{zi}(t)$ and supply temperature $T_{si}$ to each zone $i$, and the varying load for each zone ${Q}_{\text{offset}}(t)$.  
\par \textit{Privacy concerns:} We study the case when in \eqref{eq:HVAC_dynamic} the input of the system and the load are zero, i.e., ${Q}_{\text{offset}}(t), {m}_z(t) = 0$. This scenario can be interpreted when the HVAC system is switched off for instance at the end of a working day. The initial temperature of the zones in this case can be \textit{privacy-sensitive} since for instance they can be used to infer presence and absence of an employee. 
\begin{figure}[ht]
	\begin{center}
		\includegraphics[width=0.45\textwidth]{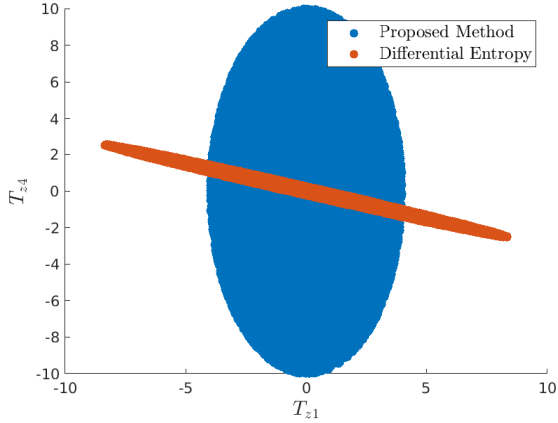}
		\caption{Confusion set \eqref{eq:ellipsoid} for the worst adversary with $\gamma =1$. We have shifted $\hat{x}_0$ to $0$ to make the comparison simpler. }\label{fig:ellips}
	\end{center}
\end{figure}
\begin{figure}[ht]
	\begin{center}
		\includegraphics[width=0.5\textwidth]{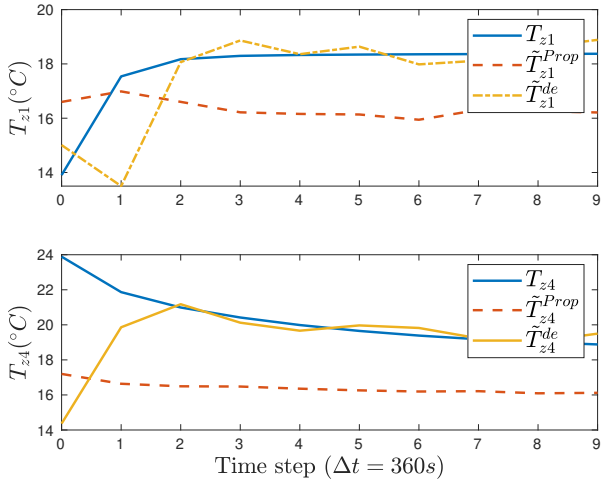}
		\caption{The true values of the measured temperatures and their private versions using the proposed method \eqref{eq:main_optimization} and \eqref{eq:opt_diff} }\label{fig:result_tem}
	\end{center}
\end{figure}
\par We consider four zones with the structure given in Figure \ref{fig:zones} and assume that temperature of the zone $1$, and $4$ are measured for monitoring reasons.
The parameters $R_{ij}$ and $(mc)_i$ are picked uniformly randomly from $[0.4,0.6]$ and $[950,1050]$, respectively, where the mean values are from the study by \cite{temp_parameter}, and Euler discretization with $\Delta t = 360$ seconds  is used to discretize the system \eqref{eq:HVAC_dynamic}.
 To compare our proposed method with the differential entropy case \eqref{eq:opt_diff},  we set the prescribed confusion set as $\Sigma_v = \diag(16, 16, 100, 100)$ in the design problem \eqref{eq:des_prob}, which basically means the  temperature in zones $3$ and $4$ are more privacy-sensitive compared to zone $1$ and $2$.
We solve \eqref{eq:main_optimization} with the given $\Sigma_v$ and obtain $\Sigma$. Next, we solve the optimization \eqref{eq:opt_diff} to find $\Sigma_{\text{de}}$ where we set $\epsilon_{p} = \tr \Sigma$ that we found for our proposed method.
\par The confusion set defined in \eqref{eq:ellipsoid} for the adversary  can be seen in Figure \ref{fig:ellips}, where we have projected the obtained hyper-ellipsoids onto $T_{z1}-T_{z4}$ plane, and the true values of the measure temperature $T_{z1}$ and $T_{z4}$ along their perturbed versions are shown in Figure \ref{fig:result_tem}.
As it can be seen by using the proposed method in this paper, the adversary's confusion set is prescribed by the system designer. On the other hand, the confusion set emerging from differential entropy while is ``larger'' in the $T_{z1}$ direction, it is ``smaller'' in the $T_{z4}$ direction which is in contrast with the desired privacy specifications concerning the privacy-wise importance of the zones. 
\section{conclusion}\label{sec:conclusion}
We have considered the problem of privacy preservation of state trajectories in data releasing dynamical systems where we have optimally designed output Gaussian noise to create a prescribed confusion set against worst case adversaries.
We have proved that a system designer can create any prescribed confusion set described by hyper-ellipsoids using correlated Gaussian noise.
 Furthermore, we have provided an approximate solution for the case of uncorrelated Gaussian noise.
   The proposed method can be pursued for preserving the privacy of the input in left-invertible linear dynamical system and also be combined with a controller design for the system. 
   \bibliography{MyReferences} 
\end{document}